\documentclass[twocolumn,showpacs,preprintnumbers,amsmath,amssymb]{revtex4}

\usepackage{graphicx}
\usepackage{dcolumn}
\usepackage{bm}

\begin{document}

\preprint{}

\title{Quantum Monte Carlo Simulation of the two-dimensional ionic Hubbard model}

\author{Bernard Martinie}
\affiliation{
D\'epartement de Physique, UFR Sciences et Techniques, Parc Grandmont, 37200 Tours, France\\
}

\date{\today}

\begin{abstract}
The Quantum Monte Carlo simulations of the ionic Hubbard model on a two-dimensional square lattice at half filling were performed. The method based on the direct-space, proposed by Suzuki and al., Hirsch and al., was used. Cycles of increasing and decreasing values of the Coulomb interaction $ U $ were performed for fixed temperature ($ kT=0.01 $). Results indicate that, at low temperature, the two insulator phases are separated by a metallic phase for weak to intermediate values of the staggered potential $ \Delta $.  For large Coulomb repulsion the system is in a Mott insulator with an antiferromagnetism order. On increasing and decreasing the Coulomb interaction $ U $ the metal-Mott insulator transition shows an hysteresis phenomenon while the metal-band insulator transition is continue. For large $ \Delta $ it seems that the metallic region shrinks to a single metallic point. However, the band insulator to the Mott insulator transition is not direct for the studied model. A phase diagram is drawn for the temperature $ kT=0.01 $. For $ \Delta =0.5 $ cycles of increasing and decreasing temperature were programmed for different values of the Coulomb interaction $ U $ . A behaviour change appears for $ U\simeq 1.75 $. This suggests that a crossover line divides the metallic region of the phase diagram.
\end{abstract}

\maketitle

\section{Introduction}
Recently some theoretical and numerical studies were published which investigate the metal-insulator transitions and the transition between the two insulator phases of the ionic Hubbard model \cite{Byczuk1,Craco1,Bouadim1,Paris1,Kancharla1}. The numerical results are obtained with the DMFT (at zero temperature) and the determinant quantum Monte Carlo method. The existence of an intermediate metallic phase between the band and the Mott insulators seems confirmed by all the authors but the nature of this phase and of the metal-insulator transitions are still under debate. 
\\ In this paper, we present results on the two-dimensional ionic Hubbard model obtained with a method based on the direct-space proposed by Suzuki and al.\cite{Suzuki1,Suzuki2} and Hirsch and al.\cite{Hirsch1,Hirsch2}. This quantum Monte Carlo method is presented in references \cite{martinie1,martinie2}. At fixed temperature, this method allows to generate  some of the most representative occupation number basis states of the model. These states are used to compute average values of energy, molar specific heat, occupancy, structure factor and rough static electric conductivity.

\section{Ionic Hubbard model}
The Hamiltonian of the ionic Hubbard model can be written
\begin{eqnarray}
\label{hamiltonian1} 
H =&&-t\sum_{\left\langle i,j\right\rangle,\sigma }\left( c^{\dagger}_{i,\sigma}c_{j,\sigma}+hc\right) +U\sum_{i}n_{i\downarrow}n_{i\uparrow}\nonumber\\&&+ \Delta \sum_{i\in A}n_{i} -\Delta \sum_{i\in B}n_{i}
\end{eqnarray}  
The square lattice is a bipartite lattice with two sublattices $ A $  and $ B $. $ c^{\dagger}_{i,\sigma} $ and $ c_{i,\sigma} $ are the fermion creation and destruction operators at the lattice site $ i $ with spin $ \sigma $. $ n_{i,\sigma}=c^{\dagger}_{i,\sigma}c_{i,\sigma} $   is the number operator. $ t $ is the hopping term between nearest-neighbor sites, $ U $ denote the on-site Coulomb repulsion, $ \Delta $ is the staggered potential between the $ A $ and $ B $ sublattices.
\section{Simulation parameters}
The square lattice contains $6\times6$ sites with periodic boundary conditions. Each elementary cell contains two sites A and two sites B. There are $ 18 $ spin up and $ 18 $ spin down (half filling). The hopping parameter $ t $ is fixed at a value $ 1 $ except for the simulations of the atomic limit where $ t=0 $. The model is decomposed in sub-systems which contain four sites. These sub-systems are grouped together in two sub-hamiltonians. The imaginary-time interval is divided into twenty slices. For each simulation five decreasing-increasing temperature cycles or decreasing-increasing interaction U cycles were programmed.There are one hundred points by curves. 
\section{The atomic limit ($ t=0 $)}
\begin{figure}
\begin{center}
\includegraphics[width=0.9\linewidth]{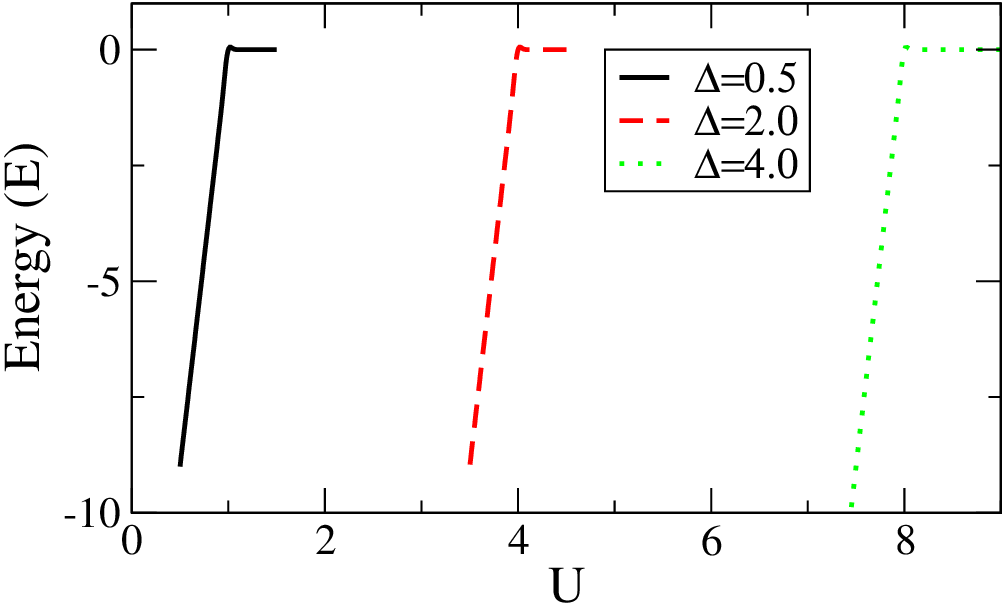} 
\end{center}
\caption{(Color online). The energy at $ kT=0.01 $ for different values of $ \Delta $.}
\label{fig:energyt0} 
\end{figure}

\begin{figure}
\begin{center}
\includegraphics[width=7cm]{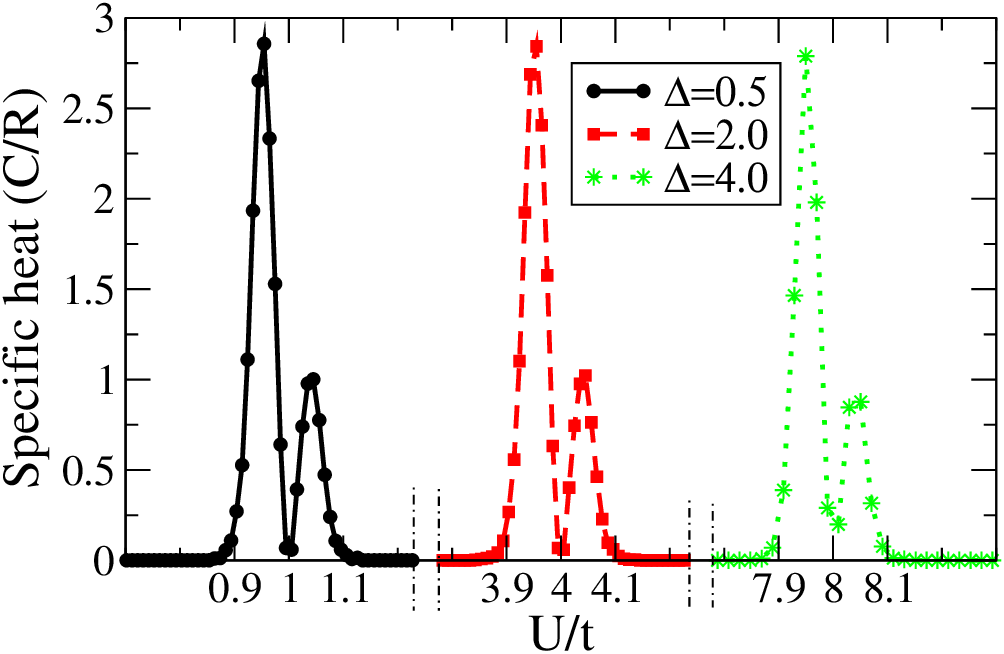} 
\end{center}
\caption{(Color online). The molar specific heat at $ kT=0.01 $ for different values of $ \Delta $.}
\label{fig:chalspect0} 
\end{figure}

\begin{figure}
\begin{center}
\includegraphics[width=0.9\linewidth]{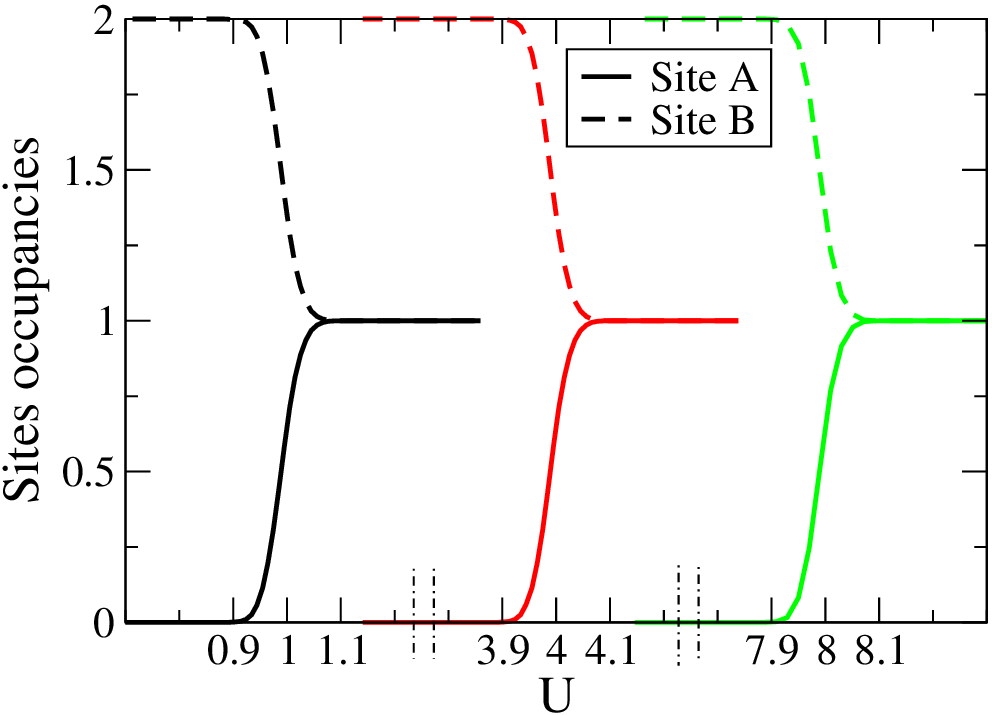} 
\end{center}
\caption{(Color online). The site occupancies at $ kT=0.01 $ for $ \Delta=0.5 $, $ \Delta=2 $ and $ \Delta=4 $.The solid lines corespond to sites A and the dashed lines correspond to sites B.}
\label{fig:occupanciest0} 
\end{figure}

Simulations were performed for the simple case of the atomic limite where $ t=0 $. For this value the Hamiltonian is diagonal, so there is no problem due to the non-commutativity, in consequence the number of slides can be one, and there is no sign problem. Cycles of increasing and decreasing values of $ U $ were programmed at the fixed temperature $ kT=0.01 $ for different values of the staggered potentiel $ \Delta $. The conductivity is always zero, so the model is an insulator whatever the values of the interactions. The Figs. \ref{fig:energyt0}, \ref{fig:chalspect0} and \ref{fig:occupanciest0} display the results.  As it is expected, one remarks that the electronic transition happens for $ U_{c}=2\Delta $ without hysteresis phenomenon. There is one spin by site for $ U\gtrsim2\Delta $ whereas only the sites of the sublattices $ B $ are occupied for $ U\lesssim2\Delta $. In this domain the energy of the model is $ E\simeq18\left( U-2\Delta\right)   $, while it is zero for $ U\gtrsim2\Delta $. There is not magnetic order.

\begin{figure}
\begin{center}
\includegraphics[width=0.9\linewidth]{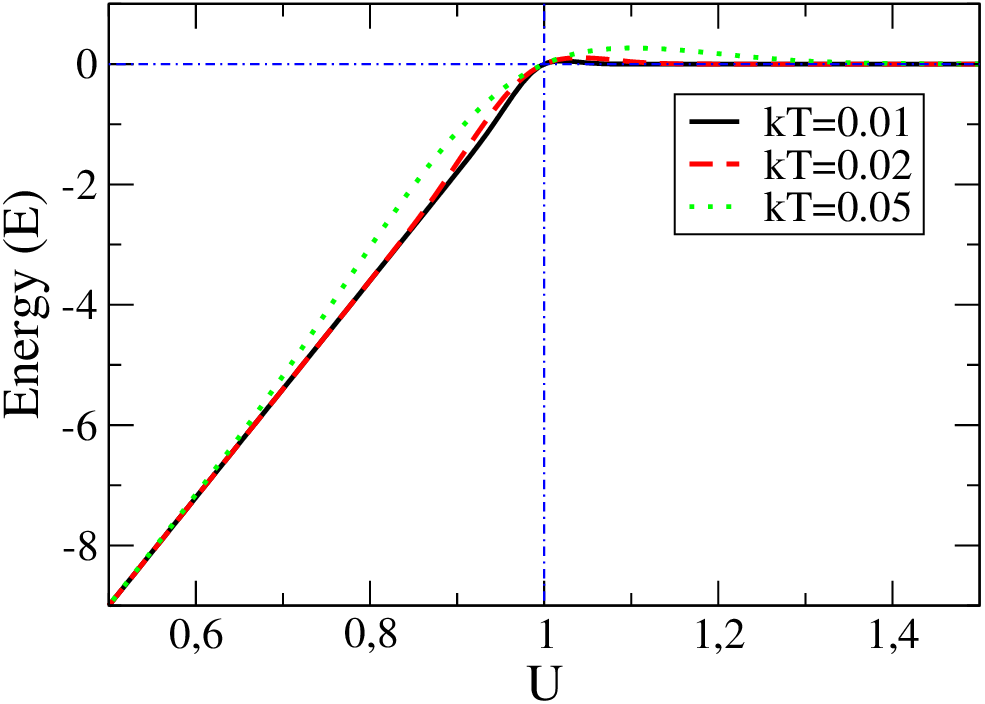} 
\end{center}
\caption{(Color online). Energy in the atomic limit $ \left( t=0, \Delta=0.5 \right)  $ for different temperatures.}
\label{fig:energy-T-t0} 
\end{figure}

\begin{figure}
\begin{center}
\includegraphics[width=0.9\linewidth]{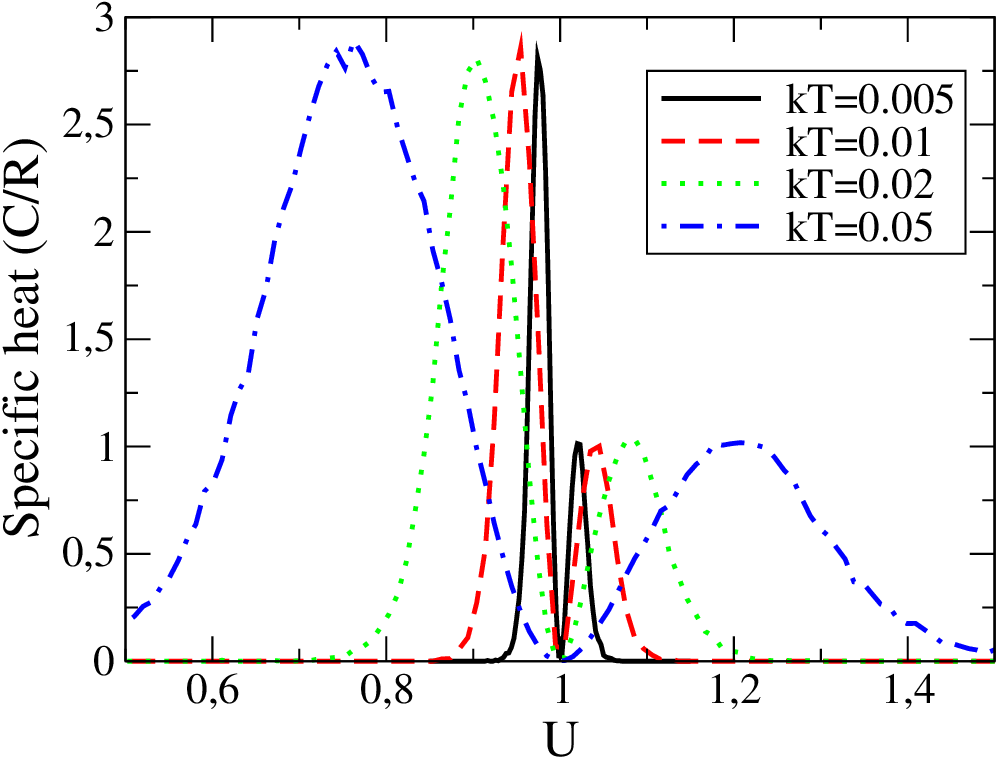} 
\end{center}
\caption{(Color online). Molar specific heat in the atomic limit $ \left( t=0, \Delta=0.5 \right)  $ for different temperatures.}
\label{fig:chalspec-T-t0} 
\end{figure}

\begin{figure}
\begin{center}
\includegraphics[width=0.9\linewidth]{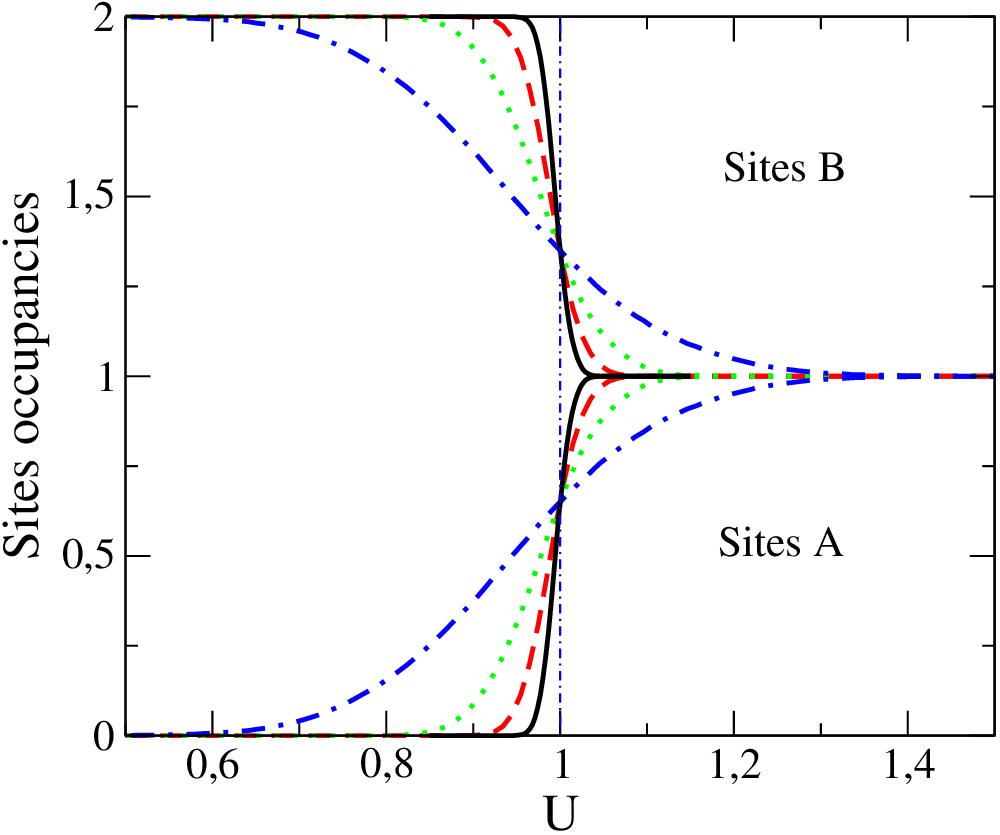} 
\end{center}
\caption{(Color online). Site occupancies in the atomic limit $ \left( t=0, \Delta=0.5 \right)  $ for different temperatures. Solid line (black) correspond to $ kT=0.005 $, the dashed line (red) correspond to $ kT=0.01 $, the dotted line (green) correspond to $ kT=0.02 $ and the dot-dash line (blue) correspond to $ kT=0.05 $.}
\label{fig:comparoccupation-T-t0} 
\end{figure}

The Figs. \ref{fig:energy-T-t0}, \ref{fig:chalspec-T-t0} and \ref{fig:comparoccupation-T-t0} show the influence of the temperature on the energy curves, the molar specific heat curves and the occupancies curves versus interaction $ U $. The occupancies curves for the different temperatures have very little error bars and cross almost exactly at $ U=2 \Delta $. This is in good agreement with the zero value of the specific heat for this value of $ U $. At this point the site A occupancy is near $ 0.66 $ while the site B occupancy is about $ 1.32 $. In this special state $ 12 $ sites A are each occupied by one spin, $ 12 $ sites B are occupied equally by one spin and $ 6 $ sites B are occupied by two spins. The energy of this state is $ E=6U-12 \Delta=0 $. Indeed, the three energy curves in Fig. \ref{fig:energy-T-t0} pass through the same point $ \left( U=1, E=0\right)  $ so $ \partial E/\partial T =0 $.\\

\begin{figure}
\begin{center}
\includegraphics[width=0.9\linewidth]{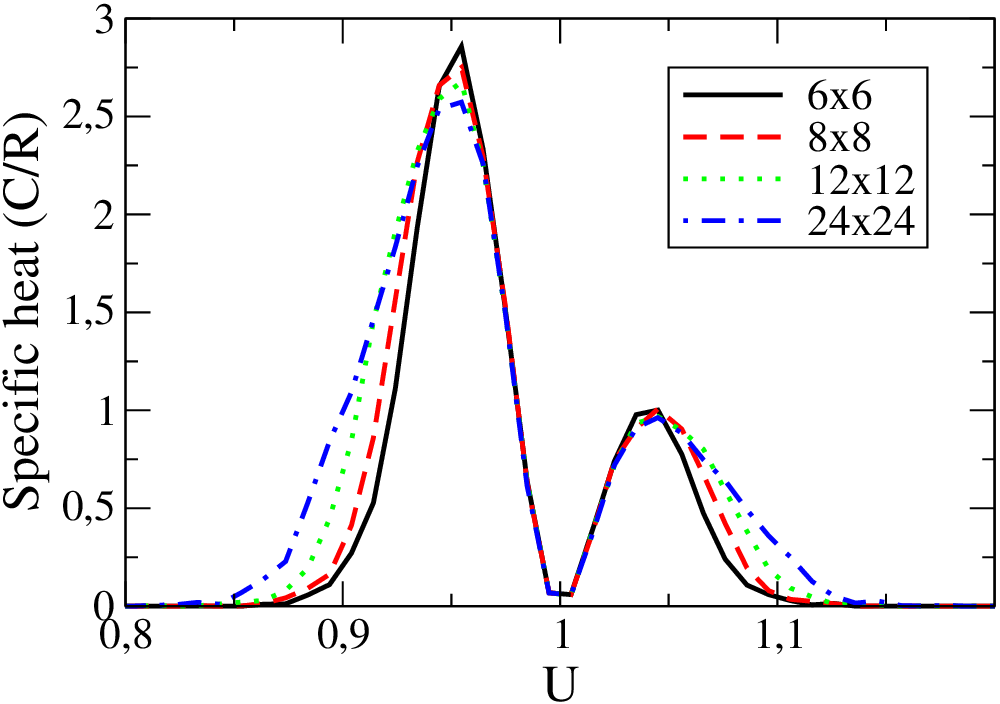} 
\end{center}
\caption{(Color online). Molar specific heat in the atomic limit $ \left( t=0, \Delta=0.5 \right)  $ for four model sizes.}
\label{fig:chalspec-N-t0} 
\end{figure}

One remarks that all the specific heat curves of the Fig. \ref{fig:chalspec-T-t0} match exactly for the abscisse $ U/kT $. The specific heat curves for four sizes of the model, at half filling, are shown in Fig. \ref{fig:chalspec-N-t0}. One remarks that these curves are similar.

\section{Model with hopping interaction $ \left( t=1 \right)  $}
\subsection{Simulations at $ T=\texttt{constant} $}
\begin{figure}
\begin{center}
\includegraphics[width=0.9\linewidth]{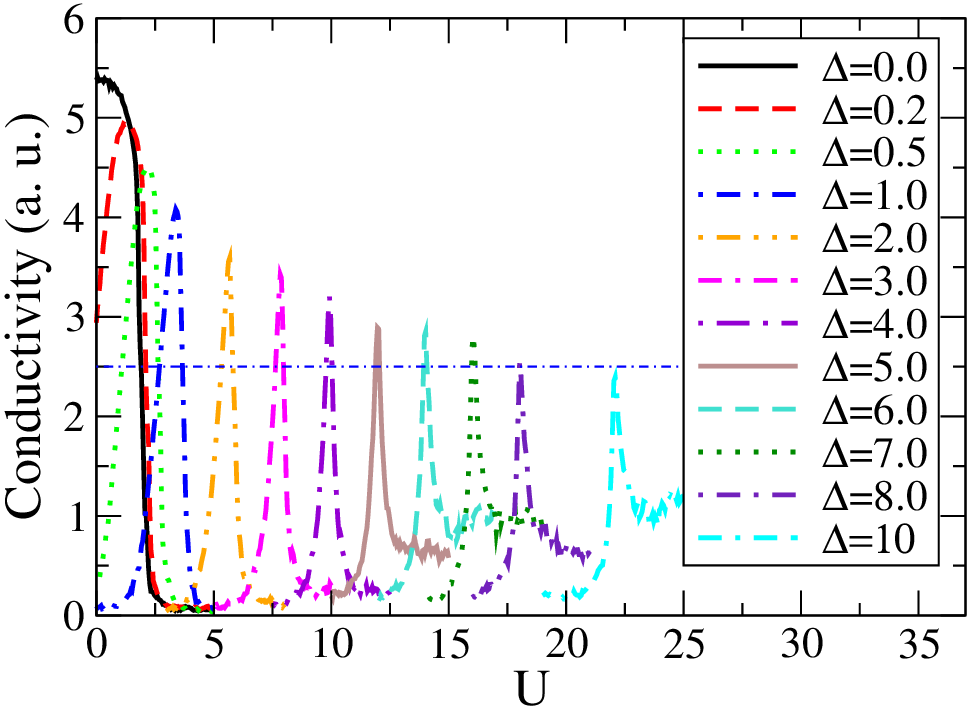} 
\end{center}
\caption{(Color online). Conductivity for low and hight values of $ U $ and different values of $ \Delta $. $ \left( t=1, kT=0.01 \right)  $ .}
\label{fig:conductivity-hight-U} 
\end{figure}

\begin{figure}
\begin{center}
\includegraphics[width=0.9\linewidth]{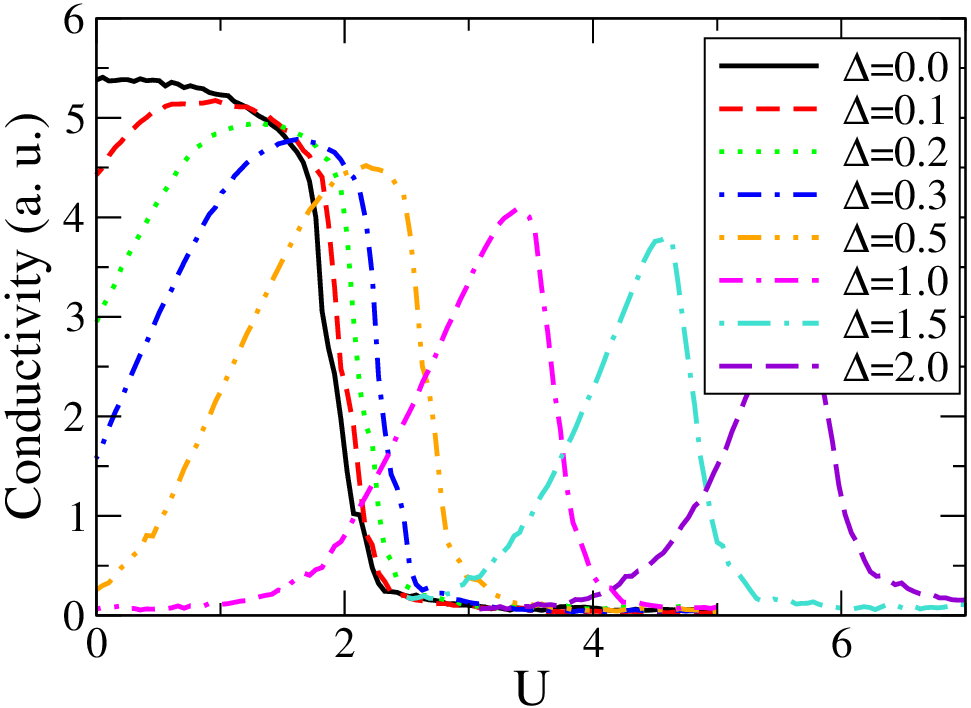} 
\end{center}
\caption{(Color online). DC Conductivity for low values of $ U $ and different values of $ \Delta $. $ \left( t=1, kT=0.01 \right)  $ .}
\label{fig:conductivity-low-U} 
\end{figure}

\begin{figure}
\begin{center}
\includegraphics[width=0.9\linewidth]{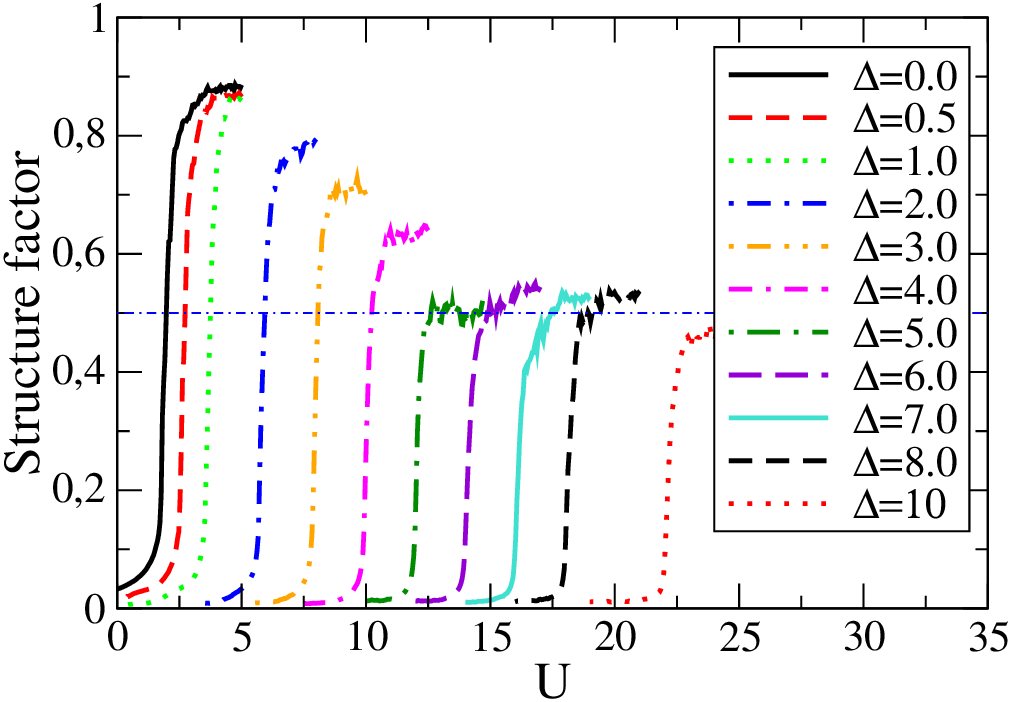} 
\end{center}
\caption{(Color online). Structure factor for different values of $ \Delta $. $ \left( t=1, kT=0.01 \right)  $ .}
\label{fig:factantiferro-U} 
\end{figure}

\begin{figure}
\begin{center}
\includegraphics[width=0.9\linewidth]{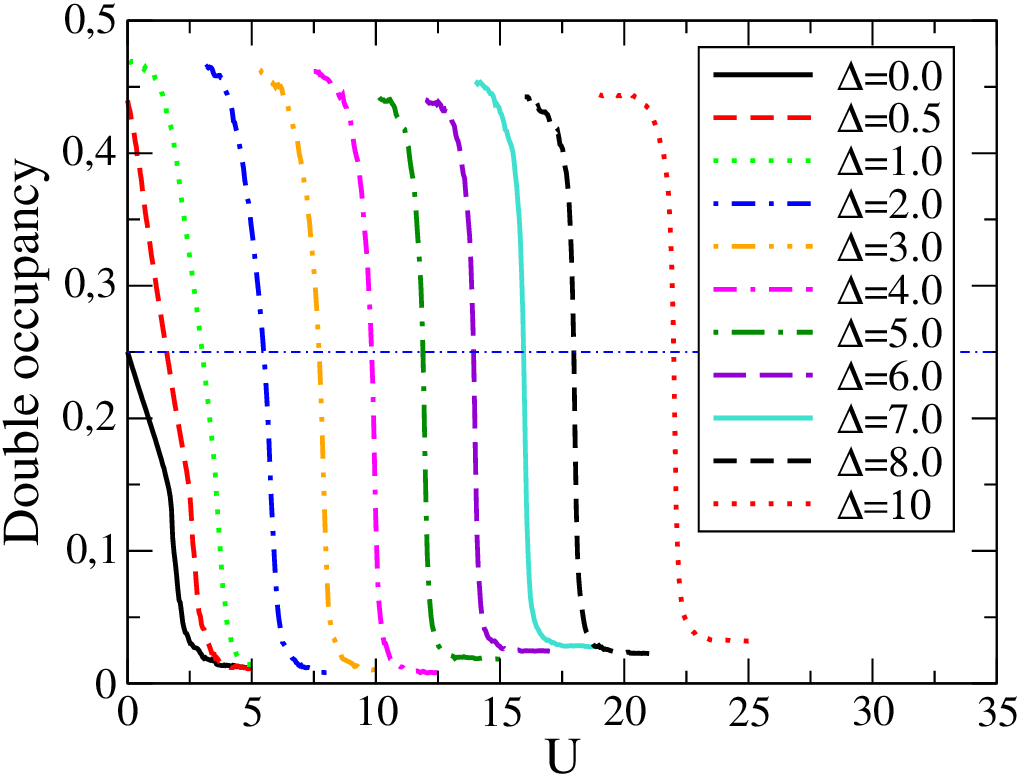} 
\end{center}
\caption{(Color online). Double occupancy for different values of $ \Delta $. $ \left( t=1, kT=0.01 \right)  $ .}
\label{fig:paire-U} 
\end{figure}

The Figs. \ref{fig:conductivity-hight-U}, \ref{fig:conductivity-low-U}, \ref{fig:factantiferro-U} and \ref{fig:paire-U} show the DC conductivity, the structure factor and the double occupancy curves for different values of the staggered potential $ \Delta $ at $ kT=0.01 $.

\begin{figure}
\begin{center}
\includegraphics[width=0.9\linewidth]{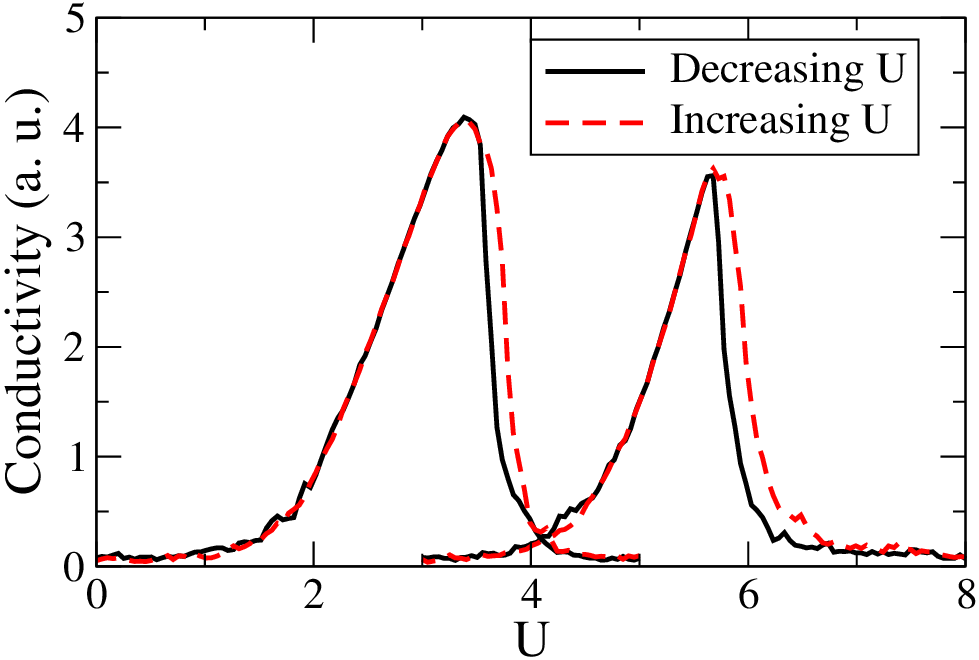} 
\end{center}
\caption{(Color online). DC conductivity for $ \Delta =1 $ and $ \Delta =2 $ for decreasing and increasing values of $ U $ at $ kT=0.01 $.}
\label{fig:conductivity-U-MD} 
\end{figure}

For $ \Delta \gtrsim 1 $ the model undergoes a transition between two insulator states. During the transition the system becomes conductor. In the low $ U $ insulator state, only the $ B $ sites are occupied by two spins. It is a band insulator (BI). All the sites are occupied by one spin in the hight $ U $ insulator state. This last insulator state presents an antiferromagnetic structure, it is a Mott insulator (MI). This transition between the two insulator states with an intermediate metallic phase was already observed \cite{Byczuk1,Craco1,Bouadim1,Paris1,Kancharla1} with other simulation methods at $ T=0 $ and $ T\neq0 $. Our results are in good agreement with those obtained within the other methods.\\
 Fig. \ref{fig:conductivity-U-MD} shows the DC conductivity for $ \Delta=1 $ and $ \Delta=2 $ for decreasing and increasing values of $ U $. An hysteresis phenomenon appears for the transition from the metal to the Mott insulator while the the curves fit for the metal-band insulator transition. This behaviour is observed for all the values of $ \Delta $. One can deduce that the MI-to-metal phase transition is a first order transition, while the BI-to-metal phase transition is continuous. This is in good agreement with the result of reference \cite{Craco1}.\\
One observes that the structure factor decreases for large $ \Delta $ while the behaviour of the double occupancy is similar for all the values of $ \Delta \gtrsim 1.0$. The caracteristcs of the BI phase ( null structure factor and double occupancy  $\approx 0.5 $) are the same for all values of $ \Delta $ whereas the insulator phase induce by increasing value of $ U $ is not a purely MI. Moreover, the conductivity of this phase is not null for $ \Delta \gtrsim 5 $. 
 
\subsection{Simulations at $ U=\texttt{constant} $}
\begin{figure}
\begin{center}
\includegraphics[width=0.9\linewidth]{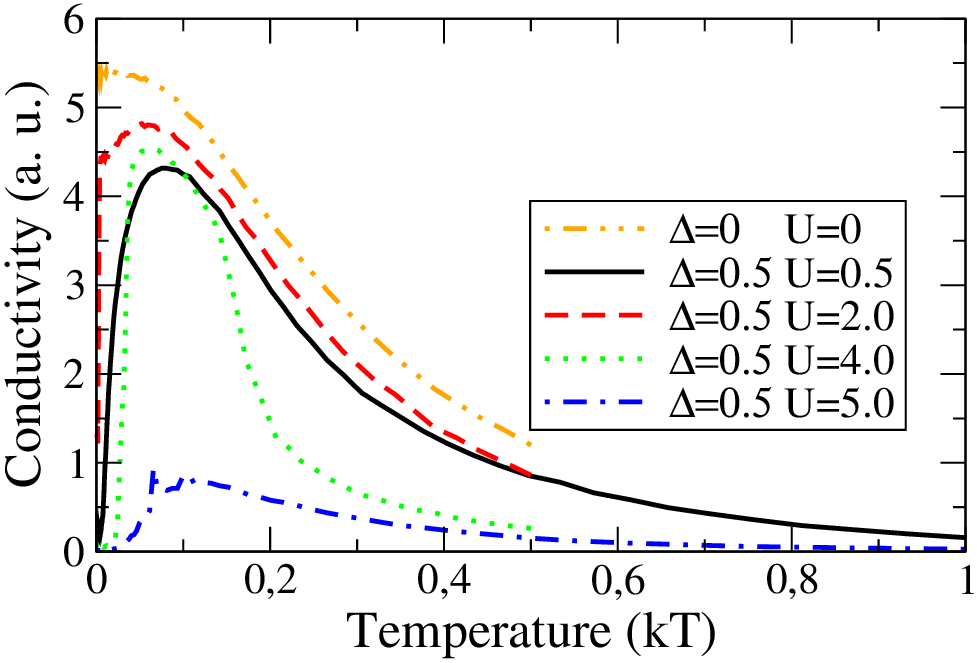} 
\end{center}
\caption{(Color online). Conductivity curves at hight temperatures for $ \Delta=0.5 $ and different values of the coulombian repulsion $ U $.}
\label{fig:conductivity-hight-T} 
\end{figure}

\begin{figure}
\begin{center}
\includegraphics[width=0.9\linewidth]{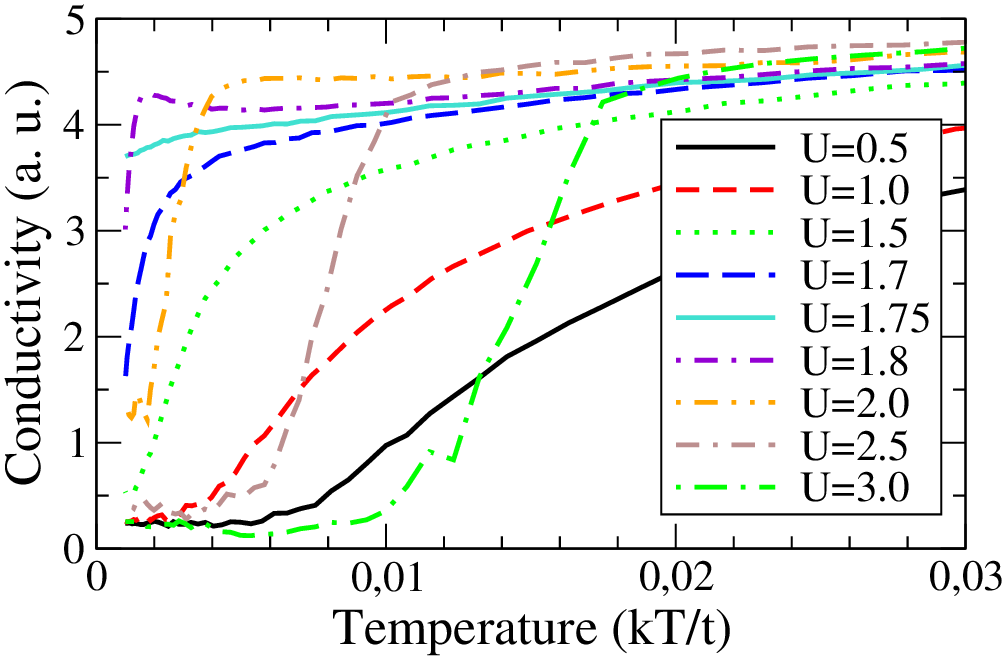} 
\end{center}
\caption{(Color online). Conductivity curves at low temperatures for $ \Delta=0.5 $ and different values of the coulombian repulsion $ U $.}
\label{fig:conductivity-low-T} 
\end{figure}

\begin{figure}
\begin{center}
\includegraphics[width=0.9\linewidth]{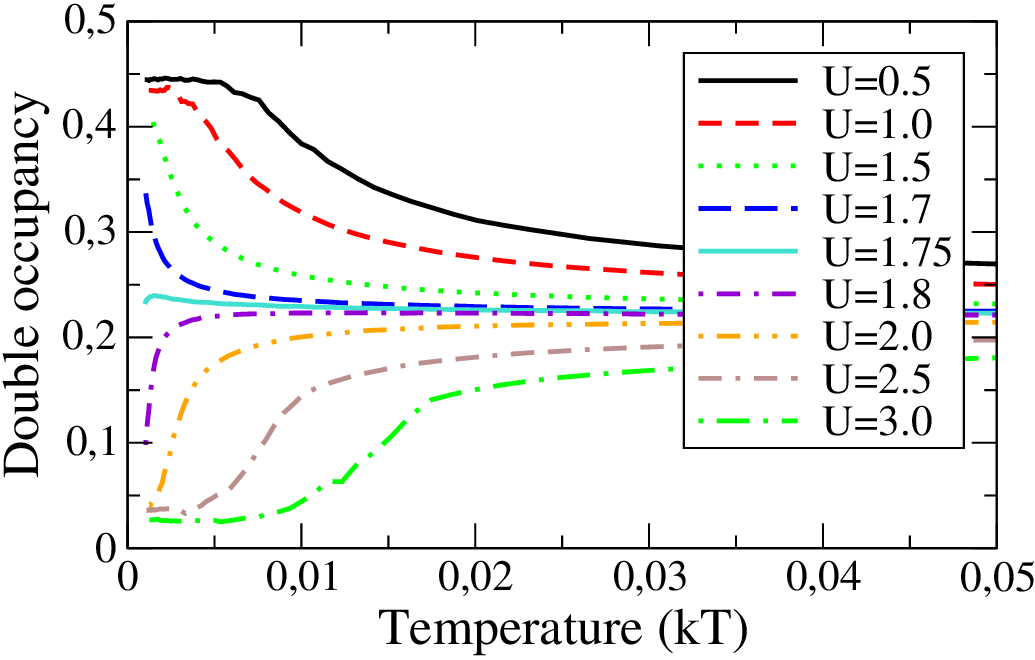} 
\end{center}
\caption{(Color online). Double occupancy curves at low temperatures for $ \Delta=0.5 $ and different values of the coulombian repulsion $ U $.}
\label{fig:doubleoccupancy-low-T} 
\end{figure}

\begin{figure}
\begin{center}
\includegraphics[width=0.9\linewidth]{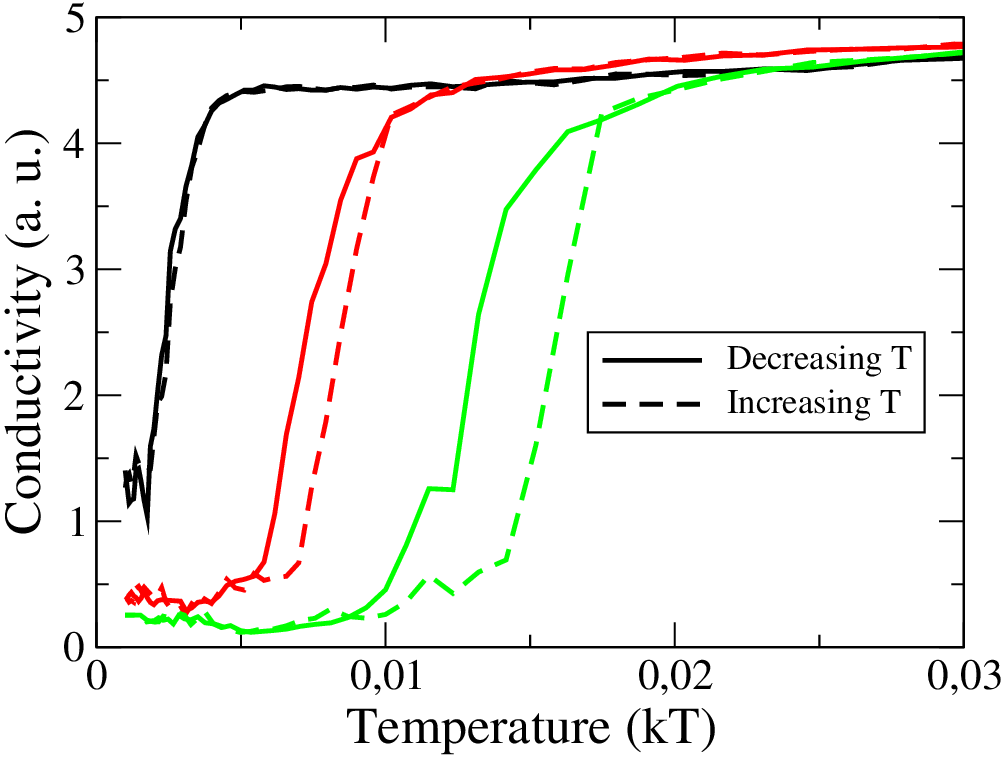} 
\end{center}
\caption{(Color online). Conductivity curves at low temperatures for increasing and decreasing temperature for $ U=2.0 $, $ U=2.5 $ and $ U=3.0 $ ($ \Delta =0.5 $). One observes an hysteresis phenomenon.}
\label{fig:hysteresis-low-T} 
\end{figure}

The Figs. \ref{fig:conductivity-hight-T} and \ref{fig:conductivity-low-T} show the conductivity curves at low and hight temperatures for $ \Delta =0.5 $ and different values of $ U $. One observes metallic behaviour for $ kT\gtrsim 0.1 $. At low temperature, for decreasing temperature (Fig \ref{fig:conductivity-low-T}), the system becomes insulator with behaviour change for $ U\simeq 1.75 $. This behaviour change can be observed equally on the double occupancy curves of Fig. \ref{fig:doubleoccupancy-low-T}. For $ U\gtrsim 1.75 $ the metal-insulator transition occurs with hysteresis phenomenon (Fig. \ref{fig:hysteresis-low-T}). One can deduce that this transition is a first order transition. On the contrary, for $ U\lesssim 1.75 $ the conductivity curves for increasing and decreasing temperature are similar.

\subsection{Phase diagram}
\begin{figure}
\begin{center}
\includegraphics[width=0.9\linewidth]{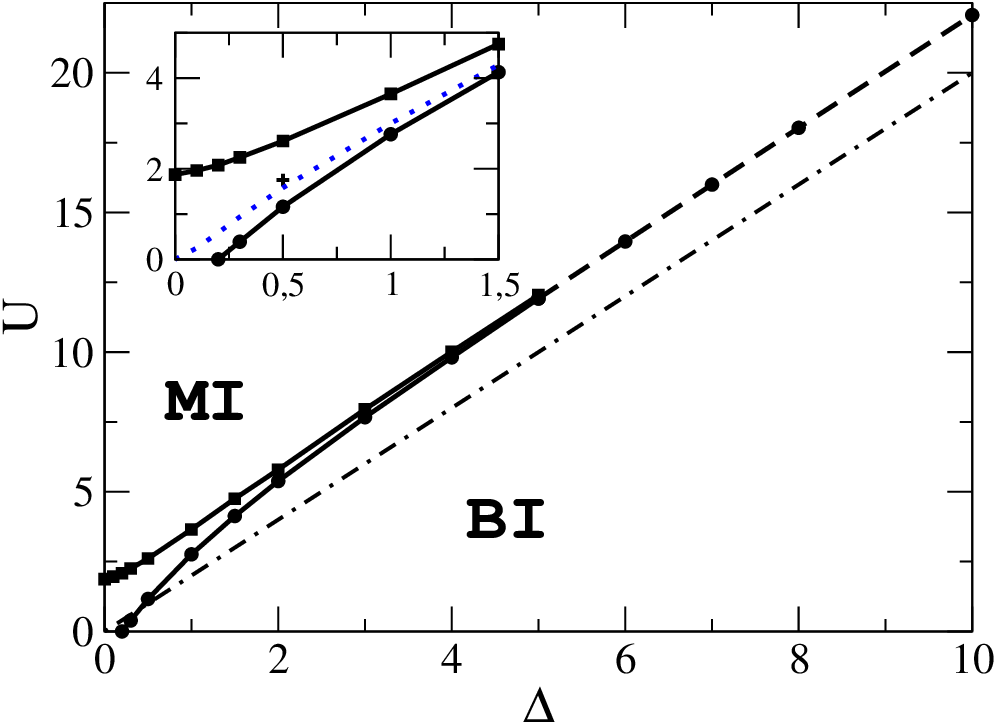} 
\end{center}
\caption{ Phase diagram of the 2D IHM at $ kT=0.01 $. The dot-dash line is the transition line at the atomic limit ($ U_{c}=2\Delta $). The symbol plus is the point at which behaviour changes. The dotted line correspond to the set of points for which double occupancy is $ 0.25 $.}
\label{fig:diagram} 
\end{figure}

The conductivity curves of Figs. \ref{fig:conductivity-hight-U} and \ref{fig:conductivity-low-U} can be used to drawn the phase diagram at the constant temperature $ kT=0.01 $. For large $ \Delta $ one can consider that the metallic region shrinks to a single metallic point. For each value of $ \Delta\lesssim 5.0 $, the coulombian interactions $ U_{c1} $ and $ U_{c2} $ which correspond with the metal-insulator transitions are determined at mid-height of the maximum conductivity. The phase diagram of the ionic Hubbard model is shown in Fig. \ref{fig:diagram}.\\
The behaviour change observed on the double occupancy curves of Fig. \ref{fig:doubleoccupancy-low-T} corresponds to the point $ \left( \Delta =0.5, U\simeq 1.75\right)  $ in the phase diagram. This suggests that a cross-over line exists in the metallic region. This line can correspond approximatly to the set of points for which the double occupancy is $ 0.25 $. For this double occupancy value $ 9 $ sites  A and $ 9 $ sites B are occupied by one spin and $ 9 $ sites B are occupied by two spins.\\
 One remarks, on the phase diagram, that the transition lines for $ t=0 $ and $ t=1 $ are parallel for large $ \Delta $. The gap between these two lines is $ \triangle U \approx 2t $.

\section{Conclusion}
\label{sec:conclusion}
Most results presented in references \cite{Byczuk1,Craco1,Bouadim1,Paris1,Kancharla1} are obtained for $ T=0 $ whereas, by principle, our simulation method works for not null temperature. However the results and the phase diagram are similar. Our results confirm the existence of a metallic region between Mott and band insulator phases. The natures of the metal-insulator transitions are different. The MI-metal transition is discontinuous while the BI-metal transistion is continuous like it is told in reference \cite{Craco1}. The metallic phase shrinks to a line for large coulombian interaction $ U $, but the BI-MI transition is not direct. Moreover, the insulator phase for $ U>U_{c} $ is not a purely Mott insulator phase. Studies with increasing and decreasing temperature show that there is a behaviour change in the metallic region which divides it into two regions. These two regions correspond to the precursor phases of MI and BI phases.



\end{document}